\documentclass[conference]{IEEEtran}
\usepackage{epstopdf}
\usepackage{multirow}
\usepackage{times,amsmath,amssymb,graphicx,epsf,psfrag,epsfig,cite,color,mathrsfs,url}
\usepackage{graphicx}
\usepackage{times}
\usepackage{amsmath}
\usepackage{graphicx, graphics,epsfig,subfigure}
\usepackage{amssymb}
\usepackage{algpseudocode}
\usepackage{algorithm}
\usepackage{dsfont}
\usepackage{latexsym}
\usepackage{tabularx}
\usepackage{comment}
\usepackage{colortbl}
\usepackage{stmaryrd}
\usepackage{amsthm}
\usepackage{amsfonts}
\usepackage{color}
\usepackage{multirow}
\usepackage{cite}
\usepackage{url}

\newcommand{\mdef}{\overset{\Delta}{=}}
\newcommand{\ie}{\emph{i.e., }}
\newcommand{\eg}{\emph{e.g., }}

\newcommand{\mi}{\mathcal{I}}

\newcommand{\mk}{\mathcal{K}}

\newcommand{\transpose}{^{\mathbb{T}}}
\newcommand{\bb}{\boldsymbol}
\IEEEaftertitletext{\vspace{-5ex}}
\begin{document}
\ifodd 1
\newcommand{\rev}[1]{{\color{black}#1}} 
\newcommand{\com}[1]{\textbf{\color{red} (COMMENT: #1)}} 
\newcommand{\response}[1]{\textbf{\color{magenta} (RESPONSE: #1)}} 
\else
\newcommand{\rev}[1]{#1}
\newcommand{\com}[1]{}
\newcommand{\response}[1]{}
\fi

\title{A Distributed Optimization Framework For Multi-channel Multi-user Small Cell Networks}

\author{Shuqin Li and Liyu Cai \\Bell Labs China, Alcatel-Lucent Shanghai Bell Co., Ltd.\\
E-mail: \{Shuqin.Li, Liyu.Cai\}@alcatel-sbell.com.cn}

\maketitle
\begin{abstract}

Small cell enchantment is emerging as the key technique for wireless network evolution.
{One challenging problem for small cell enhancement is how to achieve high data rate with as-low-as-possible control and computation overheads. As a solution, we propose a low-complexity distributed optimization framework in this paper.}
Our solution includes two parts. One is a novel implicit information exchange mechanism that enables channel-aware opportunistic scheduling and resource allocation among links. The other is the sub-gradient based algorithm with a polynomial-time complexity. What is more, for large scale systems, we design an improved distributed algorithm based on insights obtained from the problem structure. This algorithm achieves a close-to-optimal performance with a much lower complexity. Our numerical evaluations validate the analytical results and show the advantage of our algorithms.
\end{abstract}
\section{Introduction}
Wireless networks are undergoing a paradigm shift with the incoming new techniques to improve the capacity, including small cells (e.g. femtocells, picocells) and D2D (device-to-device) communications.
It is likely that networks in a near future will become so unstructured that the traditional fully centralized control will be too complicated to be efficient.
As a promising solution, a layering structure is recommended for heterogeneous networks according to the separation of data plane and control plane: the small cell layer, designed for high data rate transmission; and the macro cell layer, designed for large coverage and coordination among small cells. This paper focuses on the design problem of the small cell layer.

Compared to the traditional cells, small cells have many differences in terms of physical range, cost and functionality. First, the small cell has short transmit range. Thus simultaneous transmissions usually lead to strong interference. Second, to save the cost, small cells have rather limited processing capacities, in terms of both control power and computational power. Third, aimed for high data rates, the control overhead in small cell should be reduced as much as possible to save resource for data transmission.
To tackle the new challenges brought along by these changes,
we propose a distributed optimization framework in this paper, as a system solution for a synchronous multi-channel multi-user small cell network.
In terms of low control and computational complexity, distributed systems always show great advantages over their centralized counterparts, not to mention they have scalability, flexibility and robustness for heterogeneous network environment, and also promise improvements in the utilization of the scarce spectrum resource. However, there are two main challenges in the practical design of distributed network algorithms.

One is related to prohibitive overheads in channel information exchange. The traditional study in this area focuses on cellular networks (see \cite{love2008overview} for a good tutorial), including one-to-one (UE to BS or BS to UE), or multiple-to-one (Multiple UEs to BS) feedbacks, which can be in either analog or digital way. To reduce the system overhead, partial information exchanges\cite{love2008overview,Huang2012sum,Huang2013analytical} are proposed to feedback information of only some (not all) channels, \eg the best $m$ channel mechanism. For a distributed system, where each link needs to transmit the channel information to all others in the network due to the lack of centralized control, the system overhead of information exchange can be very high even with the partial information exchange. To overcome this challenge, we design a novel \emph{inexplicit} information exchange mechanism in our framework. It enables the channel-aware opportunistic scheduling and resource allocation by exploring the broadcast nature of wireless transmission: it encodes each link's channel information through the energy-levels of two simple signals, and enables links receiving these signals to inform the channel condition of others through simple calculations. This mechanism greatly reduces the system overhead.

The other challenge lies in rate maximization problem itself. Shown to be NP-hard\cite{luo2008dynamic} due to its non-convexity nature, this problem is extremely hard to solve even in a centralized way. To obtain the distributed solution, one popular technique is to reformulate this sum-rate-maximization problem as a non-cooperative game. Iterative Water Filling Algorithm (IWFA)\cite{yu2002distributed} has been proposed to compute the Nash equilibrium of the game, and significant efforts \cite{yu2002distributed,scutari2008optimal, luo2006analysis} have been made to establish conditions to guarantee the IWFA convergence, the existence and the uniqueness of Nash equilibrium. However, in our study, we find that IWFA is not a favorable solution for small cell networks, mainly due to its unsatisfied performance under strong interference environment, not to mention its high complexity of iterative computation and insufficient resource allocation before the convergence. After all, a Nash equilibrium solution does not necessarily mean that it can be anywhere near the social optimality, e.g. the famous  `prison dilemma'.
In contrast, we relax the rate maximization problem based on orthogonal resource allocation. We find that the concurrent transmissions in the small cell usually lead to strong interferences that greatly deteriorate the performance. Instead, the orthogonal resource allocation can achieve an optimal or near optimal throughput (shown by Hayashi and Luo in recent paper \cite{hayashi2009spectrum}). In our framework, we design distributed scheduling and power allocation algorithms with low complexity, which achieves high data rates and supports QoS and fairness among multi-users.

\section{Problem Formulation}
We consider a general multi-channel multi-user network with $\mathcal{I}\mdef\{1,2,\dots,I\}$ links\footnote{A link is composed by a pair of transmitter and its receiver, which can be a D2D pair with two User Equipments (UE), or an uplink pair between a UE and the evolved Node B (eNB).} sharing a common spectrum which is divided into $\mathcal{K}\mdef\{1,2,\dots,K\}$ frequency tones. 
With a little abuse of notation, we use the same index of link to denote its corresponding transmitter and receiver. Thus $G_{ij}^k$ denotes the channel coefficient\footnote{If we define the received signal as $y^k_i = H^k_{ii} x^k_{ii} + z_i$, then $G^k_{ii}=(H^k_{ii})^2$, and $N_0$ is the noise $z_i$'s variance.} between link $i$'s transmitter and link $j$'s receiver in tone $k$, and $N_0$ is the noise variance. The transmit power of link~$i$ in tone~$k$ is denoted as $p_i^k$, and $r_i$ denotes link~$i$'s transmit rate, which is the sum of link~$i$'s rates in all tones.
%
%
%
Therefore, the weighted sum Rate Maximization (RM) problem for the system can be formulated as follows:
\begin{eqnarray}
\text{RM}: &\underset{p_i^k\ge 0}{\text{Maximize}}&\sum_{i\in \mathcal{I}} \theta_ir_i \nonumber\\
&\text{Subject to}& \sum_{k\in\mathcal{K}}p_i^k\le P_{i0},\;\forall\, i\in\mi \label{eq:power_con}
\end{eqnarray}
where $\theta_i$ is the weight of each link~$i$, a parameter predetermined by the system to enable some design flexibility (\eg priorities, fairness, and QoS requirements etc.). The constraint in \eqref{eq:power_con} denotes that each link~$i$ has a limited transmit power of each link, \ie no larger than its maximum power $P_{i0}$.

We apply orthogonal resource allocation (\ie tones are orthogonally shared by links) in our solution. Thus the transmit rate of link~$i$ can be calculated according to Shannon formulation $r_i\mdef\sum_{k\in\mathcal{K}}\log\left(1+{g_{i}^kp_i^k}\right)$, where $g_i^k\mdef\frac{G_{ii}^k}{N_0}$ is the normalized channel gain for link $i$ in tone~$k$.


There are two technical challenges to  solve the RM~problem in a distributed manner. First, the channel gain $g_i^k$ is private information of each link. Thus a simple and effective information exchange mechanism is needed.
Second, the RM~problem has been shown to be NP-hard\cite{luo2008dynamic}. Thus an efficient algorithm is needed to compute an approximated solution with satisfactory performances.

\section{The Optimization Framework}
\begin{figure}[htb]
\centering
\includegraphics[scale=0.36]{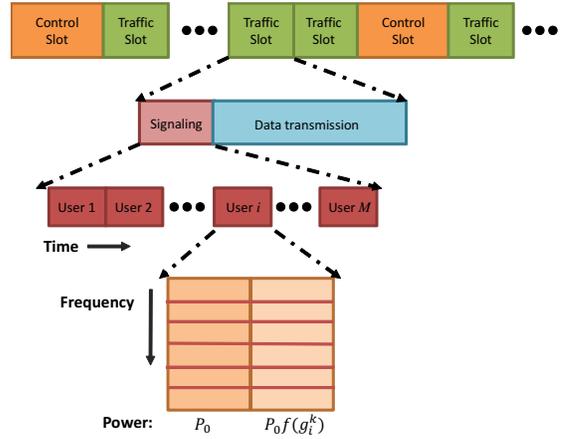}
\caption{System Timeline: each traffic slot is composed by a signaling slot and a transmission slot, and each signaling slot is further divided into $M$ sub-slots. Each link will be allocated such a sub-slot to do two signal transmissions.}
\label{fig:timeline}
\end{figure}

We propose a distributed optimization framework to tackle the aforementioned challenges. 
This framework includes two key functional modules: one \emph{signaling} module that enables channel information exchange between links, and the other \emph{scheduling and resource allocation} module that solves the RM~problem.
Operations of these functional modules in the framework will run in a time-slot base. The timeline of the framework is illustrated in Fig.~\ref{fig:timeline}. In the section~\ref{sec:signaling} and ~\ref{sec:scheduling}, we will discuss the design of each module in more details, respectively.

\subsection{Inexplicit Signaling}
\label{sec:signaling}
In the signaling module, we design an inexplicit information exchange mechanism that enables links to coordinate in scheduling and resource allocation to achieve a good system performance in a distributed manner.
Specifically, this mechanism encodes the channel condition of each link through the energy-levels of two simple signals, and enables each link to inform the channel condition of others through simple calculations. It works in the signaling slot of each traffic slot as shown in Fig.~\ref{fig:timeline}. Each signaling slot is further divided into $M$ sub-slots, where $M\ge I$ guarantees that each link in the network can be allocated one sub-slot. The allocation can follow links' predetermined indices (\eg according to their priorities).
In the signaling slot, links sequentially inform their channel gain in their own sub-slot as following two steps:
\begin{itemize}
    \item Step 1: each link~$i$ broadcasts a signal with power $P_0$ in each tone $k$.
	 \item Step 2: each link~$i$ broadcasts a signal with power $P_0 f(g_i^k)$ in each tone $k$.
\end{itemize}
$f(g_i^k)$ is public information of links, a function predetermined by the system, which can be any monotonically increasing function of the channel gain $g_i^k$. For example, a nice choice of $f(g_i^k)$ can be the cumulative distribution function (CDF) of the channel gain $g_i^k$.
When another link, say a link~$j$, hears these two signals, denoted as $S_1$ and $S_2$,  it can quickly obtain the channel gain~$g_i^k$ by the following simple calculation:
\begin{equation}
\label{eq:signaling}
    g_i^k = f^{-1}\left(\frac{S_2}{S_1}\right)= f^{-1}\left(\frac{g_{ij}^kP_0 f(g_i^k)}{g_{ij}^kP_0 }\right)
\end{equation}

It is more convenient to adopt discrete function of $f(g_i^k)$ for implementation, since each $g_i^k$ is also a sampled and quantified discrete value in practical systems. Thus the values of the function $f(g_i^k)$ are stored as a table, and the operation in \eqref{eq:signaling} is simply a table searching.

We can take the following simple example to see this process in a more clear way. Suppose that the channel gains are quantified as three values: HIGH, MIDDLE, LOW, and the system can simply predetermine
\begin{equation}
    f(g_i^k)=
\begin{cases}
1 & \text{if }g_i^k=\text{HIGH},\\
\frac{2}{3}& \text{if } g_i^k=\text{MIDDLE},\\
\frac{1}{3}& \text{if } g_i^k=\text{LOW}.
\end{cases}
\end{equation}
Next, when the link~$j$ obtains the power quotient of two received signals, say, $2/3$, then it immediately infers that the channel gain of $g_i^k$ is the MIDDLE value.

In practical implementation, we can choose much denser samplings for quantifying channel conditions $g_i^k$ (\eg far more than three channel conditions in this toy example). Thus the more accurate channel conditions, the more benefits for computing a better scheduling and resource allocation in the transmission module.

\subsection{Scheduling and Resource Allocation}
\label{sec:scheduling}
After the signaling slot, each link knows the channel information of all other links. The transmission module of each link will individually calculate the scheduling and resource allocation solution to the RM problem in the next data transmission slot as shown in Fig.~\ref{fig:timeline}. We first propose a sub-gradient algorithm which converges to the optimal value guaranteed by strict mathematical proofs. Based on the insights obtained by this algorithm, we then proposed a more efficient sub-optimal solution with much lower complexity, and yet a close-to-optimal performance.

\subsubsection{Optimal Subgradient Algorithm}
\label{sec:opt-algorithm}
The RM problem is shown to be NP-hard. To see it clearly, we introduce binary variables $T_i^k$: tone~$k$ is assigned to link~$i$, if $T_i^k=1$; otherwise $T_i^k=0$. The transmission rate is
$   r_i\mdef\sum_{k\in\mathcal{K}}T_i^k\log\left(1+\frac{g_{ii}^kp_i^k}{T_i^k}\right).
$
Since the introduction of $T_i^k$ does not change the value of transmission rate, the resulting problem is equivalent to the RM problem. We can see that the NP-hardness of the RM problem roots in the combinatorial choice of channel assignment, \ie the binary $T_i^k$.
%
%
To resolve this difficulty, we relax the binary variable $T_i^k$ into the range from 0 to 1, \ie $T_i^k\in [0,1]$. The corresponding physical meaning is that each channel can be Time Sharing (TS) among several links:
\begin{eqnarray}
\text{TS:}&\underset{p_i^k\ge 0,\; 0\le T_i^k\le 1}{\text{Maximize}}&\sum_{i\in \mathcal{I}} \theta_ir_i \nonumber\\
&\text{Subject to}& \sum_{k\in\mathcal{K}}p_i^k\le P_{i0},\;\forall\, i\in\mi \label{eq:power_con_ts}\\
&& \sum_{i\in\mathcal{I}}T_i^k\le 1,\;\forall\, k\in\mk \label{eq:time_con}
\end{eqnarray}
We can show that the TS problem is a convex optimization problem, and that there is no duality gap.  Thus we can solve the equivalent dual problem by subgradient algorithms\cite{Bertsekas1999nonlinear}. Due to the page limit, technical details are provided in Appendix A and B of our technical report.


\subsubsection{Low Complexity Suboptimal Algorithm}
\label{sec:sub-algorithm}
There are several major difficulties to apply the subgradient algorithm in the practical system. First, despite its polynomial complexity, the subgradient search still converges too slowly to be useful for the scheduling on a fast scale. Second, the time-sharing result will cause overhead of synchronizing links to share data transmission slot together. However,  by using the structure information revealed by the subgradient algorithm, we propose the following more efficient Sub-Optimal Algorithm (SOA).

In the subgradient solution, given the optimal dual variable $\boldsymbol{\lambda}^*$, the channel assignment is determined by sorting the links on each tone according to a complex metric $\xi_{i}^k$, and then given the optimal channel assignment, the power allocation is given by a water filling solution (refer to Appendix B of our technical report). In the SOA, we adopt the same two phases, but modify them to reduce the complexity.

{\bf Channel Assignment Phase:} In this phase, each channel will be assigned to at most one link. Instead of $\xi_{i}^k$, we introduce a new metric \eqref{eq:marginal_rate} based on the equal power allocation over all tones assigned to a link. The pseudo code is given in Algorithm~\ref{alg:greedy_algorithm}.
\begin{algorithm}[htb]
\caption{Channel Assignment for SOA}
\label{alg:greedy_algorithm}
\begin{algorithmic}[1]
\State Initialization: \emph{No-Assigned-Channel}=0,
\Statex \quad\quad\quad\quad\quad\;\;\emph{Assigned-Channel-Set}$_i=\emptyset$
\While{\emph{No-Assigned-Channel} $<K$}
	\State Update the index \emph{Channel-To-Assign}$_i$ for each link~$i$
	\State Update the metric \emph{Marginal-Rate}$_i$ for each link~$i$
	\State Find $i^*\gets \arg\max $\{\emph{Marginal-Rate}$_i>0,i\in\mi$\}
	\State Assign the channel \emph{Channel-To-Assign}$_{i^*}$ to link~$i^*$ by updating:
   \Statex \quad\quad\;\;\emph{No-Assigned-Channel} $\gets$ \emph{No-Assigned-Channel} $+1$
	\Statex \quad\quad\; \emph{Assigned-Channel-Set}$_{i^*}$ $\gets$ \emph{Assigned-Channel-Set}$_{i^*}$ $\cup$ \emph{Channel-To-Assign}$_{i^*}$
\EndWhile
\end{algorithmic}
\end{algorithm}
In this algorithm, \emph{No-Assigned-Channel} is a counter denoting how many tones have been assigned, and \emph{Assigned-Channel-Set}$_i$ denotes the tones assigned to link~$i$. The index \emph{Channel-To-Assign}$_i$ denotes the tone to be assigned to the link~$i$ in each iteration. It is determined as follows: we sort the tones based on the
channel gain for each individual link. For each link $i$, \emph{Channel-To-Assign}$_i$ is the tone index with the largest channel gain in  the unassigned channels. The metric \emph{Marginal-Rate}$_i$ is defined as the marginal rate if the channel \emph{Channel-To-Assign}$_i$ is allocated to the link, \ie
\begin{align}
\label{eq:marginal_rate}
\text{\emph{Marginal-Rate}}_i&=\theta_i\sum_{k\in\text{\emph{ACS}}_i\cup \text{\emph{CTA}}_i}\log\left(1+\frac{P_{i0}g_{i}^k}{|\text{\emph{ACS}}_i|+1}\right) \nonumber\\
&-\theta_i\sum_{k\in\text{\emph{ACS}}_i}\log\left(1+\frac{P_{i0}g_{i}^k}{|\text{\emph{ACS}}_i|}\right)
\end{align}
where \emph{ACS}$_i$ is the abbreviation of \emph{Assigned-Channel-Set}$_i$, and \emph{CTA}$_i$ is the abbreviation of  \emph{Channel-To-Assign}$_i$.

 {\bf Power Allocation Phase:} After the channel assignment, we can adopt equal power allocation, which removes all computations of the water filling solution. However, in the simulation, we find that it can maintain almost the same performance of water filling solution.

It is not hard to see that the total complexity of SOA is $\mathcal{O}(I K \log(K))$.

\section{Performance}
\label{sec:performance}
The performance of the proposed framework is examined in this section. Several different indoor small cell scenarios (urban or suburban, inside or outside door, and also different radius) are considered and tested. Due to the page limit, we only show results for one typical scenario. For other scenarios and discussions on implementation issues, readers can refer to Appendix C and D of our technical report.

Here we consider a urban indoor small cell scenario, where different number (2--10) of links are uniformly distributed (\ie both transmitters and receivers are randomly dropped) in a small cell with radius of 25m, sharing 10 different tones, with 180kHz for each tone\footnote{A typical system would be the 20MHz LTE system: 10 resource blocks, 180kHz for each, with 2MHz for guard band.}. The pathloss model is $\text{PL(dB)}=38.46+20\log_{10}R+0.7d_\text{2D,indoor}+18.3n^{(n+2)/(n+1)-0.46}+q\,L_{iw}$ with $\text{lnH}(\text{LOS})$ fading\footnote{
Parameter $d_\text{2D,indoor}$ is the distance insider the room. $n$ is the number of penetrated floors. $q$ is the number of walls, $L_{iw}$ is the penetration loss, 5dB. Readers can refer to 3GPP's standardization 36.814 for more explanations. The result in this letter use the settings $d_\text{2D,indoor}=25m$ $n=0$, $q=1$. We test other values of these parameters, the value of throughputs will be different, but the trends of curves are similar.}. The maximum power of each link is $P_{i0}=20\,\text{dBm}, \forall i\in \mi$.
The noise power density is $N_0=-174\text{dBm}/\text{Hz}$. 

\begin{figure}[htb]
\centering
\includegraphics[scale=0.45]{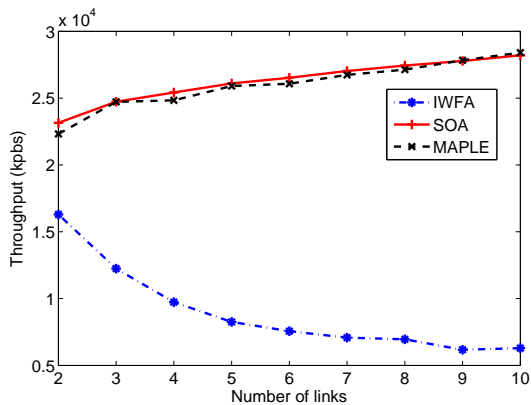}
\caption{Performance comparisons of three algorithms}
\label{fig:simulation}
\end{figure}

Two popular algorithms are compared with our algorithm (SOA). One is the  distributed algorithm IWFA. The weights are set as $\theta_i=1,$ $i\in\mi$, so that IWFA can work. The other is MAPLE\cite{qian2009maple}, which is a centralized algorithm, proved to be throughput-optimal for single channel scenario. In our simulation, we extend it for the multi-channel scenario by averagely splitting the link power budget, independently running MAPLE in each tone, and taking the sum rate of all tones as its throughput. Fig.~\ref{fig:simulation} shows the throughputs of these three algorithms as the number of links increases, in which each point in each curve is with 100 trials. Tab.~\ref{tab:time} records the average execution time of one trial for three algorithms\footnote{All trails are run in exactly the same computational environment. In Tab.~\ref{tab:time}, we use the following abbreviations for MAPLE's average execution time: $1.80e3=1.80\times 10^3$, $7.71e3=7.71\times 10^3$ and $3.24e4=3.24\times 10^4$.}.

As shown in Fig.~\ref{fig:simulation}, the proposed SOA has a very similar (sometimes even better) throughput as MAPLE. Their performances are increasing with the number of links increasing, which  is mainly due to the multi-use diversities in the frequency domain. We observe that the performance of IWFA deteriorates quickly as the number of links increases. Compared with it, SOA has a large performance gain, from 142\% for 2 links to 449\% for 10 links. As we know, spatial diversity gain and interference are two sides of one coin for concurrent transmissions. Designed to explore the spatial diversity, IWFA tries to enable simultaneous transmissions as many as possible, which is in general not suitable for such a strong interference environment as small cells as shown in our simulation, for the performance loss caused by interference greatly overweighs the gain possibly brought by spatial diversity. 

\begin{table}[tb]
\centering
\begin{tabular}{|c|c|c|c|c|c|c|c|c|c|c|}
\hline
No. of link& 2&3&4&5&6&7&8&9&10\\
\hline
SOA ($\mu$s)&32.6&32.3&31.1&29.7&29.4&29.9&29.5&30.1&29.6\\
\hline
IWFA (ms)&0.36&0.59&0.90&1.27&1.72&2.21&2.79&3.61&4.26\\
\hline
MAPLE (s)&0.06&0.22&0.75&2.56&11.5&55.8&1.80e3&7.71e3&3.24e4\\
\hline
\end{tabular}
\caption{Execution Time of Three Algorithm}
\label{tab:time}
\end{table}

In terms of system complexity of algorithms, there are two parts, control complexity and computational complexity. For the first part, it highly relates to the concrete system implementation. However, several things are for sure: First, the inexplicit signaling mechanism proposed in Sec.~\ref{sec:signaling} has great advantage for reducing feedback overheads, and can be customized to work with other resource allocation algorithms. Second, enabling by the inexplicit signaling, SOA can independently work on each link, and obtain the resource allocation result by one calculation as the centralized algorithms, \eg MAPLE. However, IWFA usually takes dozens of iterations (\ie one calculation and resource allocation per link per iteration) before its convergence. For the computational complexity, the comparisons will be more direct, just by checking the execution time in Tab.~\ref{tab:time}, which is proportional to the algorithm's computational complexity. It is clear that SOA has a very low complexity.


\section{Conclusion}
\label{sec:conclusion}

In this paper, we propose a distributed optimization framework for the
multi-channel multi-link small cell network, which enables inexplicit information exchange and channel-aware opportunistic scheduling and resource allocation. It achieves a close-to-optimal performance with very low complexity, and thus shows great potentials for small cell applications in future heterogeneous networks.

\bibliographystyle{IEEEtran}

\appendix

\subsection{Derivation of Sub-gradient Algorithm}
The dual problem of the TS problem is shown as follows:
\begin{equation}
\label{eq:dual_fun}
    \underset{(\bb{\mu},\bb{\lambda})\ge 0}{\text{Minimize}}\;\; D(\bb{\mu},\bb{\lambda}),
\end{equation}
where $\lambda_i$ and $\mu^k$, $i\in\mi$, $k\in\mk$ denote the Lagrange multipliers associated with constraints \eqref{eq:power_con_ts} and \eqref{eq:time_con}, and the bold font denotes the vector form of the corresponding variable.  The dual function is defined as
\begin{equation}
    D(\bb{\mu},\bb{\lambda})=\max_{\bb{p}\ge 0, \bb{T}\in [0,1]}L(\bb{\mu},\bb{\lambda};\bb{p},\bb{T}).
\end{equation}
and the Lagrangian is defined as
\begin{align*}
L(\bb{\mu},\bb{\lambda};\bb{p},\bb{T})\!=\!\!\!\sum_{i\in\mi}\!\theta_ir_i \!-\!\!\!\sum_{k\in\mk}\!\!\mu^k\!\!\left(\!\sum_{i\in\mi}T_i^k\!-\!1\!\!\right)\!\!-\!\!\sum_{i\in\mi}\!\!\lambda_i\!\!\left(\!\sum_{k\in\mk}p_i^k\!-\!\!P_{i0}\!\!\right)\!.
\end{align*}

We first calculate the dual function as follows. For given $\bb{\mu},\bb{\lambda}$, we obtain the optimal $\bb{p}^*$ by maximizing the Lagrangian $L(\bb{\mu},\bb{\lambda};\bb{p},\bb{T})$ through the first-order derivative condition:
\begin{equation}
\label{eq:opt_p}    p^{k*}_i=\frac{T_i^k}{h_{i}^k}\left[\frac{\theta_ih_{i}^k}{\lambda_i}-1\right]^+, \;\forall,i\in\mi,k\in\mk,
\end{equation}
where we use the notation $(\cdot)^+\mdef \max(\cdot,0)$. With the optimal $\bb{p}^*$, the resulting Lagrangian is
%
\begin{align*}
    L(\bb{\mu},\bb{\lambda};\bb{p}^*,\bb{T})\!=\!\!\!\sum_{i\in \mi}\!\sum_{k\in\mk}\!\! \left(\xi_i^k({\lambda_i})\!-\!\mu^k\right)\!T_i^k
\!+\!\!\!\sum_{k\in\mk}\mu^k\!\!+\!\!\!\sum_{i\in\mi}\!\lambda_iP_{i0},
\end{align*}
where
\begin{align}
\xi_i^k({\lambda_i})\!\mdef\!
\begin{cases}
\theta_i\left[\log\left(\frac{\theta_ih_{ii}^k}{\lambda_i}\right)-\frac{\lambda_i}{\theta_ih_{i}^k}+1\right]& \text{If } \theta_ih_{i}^k>\lambda_i,\\
0 &\text{Otherwise. }
\end{cases}
\end{align}
It is not hard to see that $L(\bb{\mu},\bb{\lambda};\bb{p}^*,\bb{T})$ is linear in $\bb{T}$. Thus we quickly obtain the optimal value $\bb{T}^*$ by maximizing $L(\bb{\mu},\bb{\lambda};\bb{p}^*,\bb{T})$:
\begin{equation}
\label{eq:opt_T}
    T_i^{k*}=
	 \begin{cases}
		1, & \text{If } \xi_i^k(\lambda_i)>\mu^k,\\
     \text{any value}\in[0,1], & \text{If } \xi_i^k(\lambda_i)=\mu^k,\\
	   0, & \text{If } \xi_i^k(\lambda_i)<\mu^k.
	 \end{cases}
\end{equation}
With the optimal $\bb{p}^*$ and $\bb{T}^*$, we obtain the dual function as
\begin{align}
\label{eq:dual_fun_3}
    D(\bb{\mu},\bb{\lambda})\!=\!\!\sum_{k\in\mk}\sum_{i\in \mi} \left(\xi_i^k({\lambda_i})-\mu^k\right)^+\!\!
+\!\!\sum_{k\in\mk}\mu^k\!\!+\!\!\sum_{i\in\mi}\lambda_iP_{i0}
\end{align}

Next we solve the dual problem \eqref{eq:dual_fun} by minimizing the dual function \eqref{eq:dual_fun_3} over $\bb{\mu}$ for given $\bb{\lambda}$.

We can show that the optimal value of $\bb{\mu}$ is given by
\begin{equation}
\label{eq:multiplier_mu}
    \mu^{k*}= \max\limits_{i\in\mi}\{\xi_i^{k}\}\mdef \xi_{(1)}^{k},k\in\mk.
\end{equation}
where $\xi_{(1)}^{k}$ denotes the largest elements of the set $\{\xi_{i}^{k}\}_{i\in\mi}$.
Substituting all above results, we have the equivalent dual problem, shown as follows:
\begin{equation}
\label{eq:dual_fun2}
    \underset{ \bb{\lambda}\ge 0}{\text{Minimize}}\;\; D(\bb{\lambda})
\end{equation}
where
$
    D(\bb{\lambda})\mdef\sum_{k\in\mk} \xi_{(1)}^k({\lambda_i})+\sum_{i\in\mi}\lambda_iP_{i0}.
$

Problem \eqref{eq:dual_fun2} is a single variable convex optimization problem.
To solve it, we adopt a subgradient-based search and update $\bb{\lambda}$ as follows:
\begin{eqnarray}  \bb{\lambda}(t+1)\!=\!\left[\bb{\lambda}(t)-\alpha(t)\left(\bb{P}_0-\sum_{k\in\mk}\bb{p}^{k*}(t)\right)\right]^+
\end{eqnarray}
where $\bb{p}^{k*}$ are given by \eqref{eq:opt_p}, $\alpha(t)$ is the step size. By choosing a diminishing step size that is \emph{square summable but not summable}, the subgradient algorithm can be proved to converge to the optimal solution within polynomial time\cite{Bertsekas1999nonlinear}. We give the proof in the next section.

\subsection{Convergence of Sub-gradient Algorithm}
\begin{proof}
Recall that a subgradient of $f$ at $x$ is any vector $v$ that satisfies the inequality $f(y)\ge f(x) + v\transpose (y-x)$ for all $y$. By this definition, it is not hard to show that
$$g\mdef \bb{P}_0-\sum_{k\in\mk}\bb{p}^{k*}$$
is the subgradient of function $D(\bb{\lambda})$ at $\bb{\lambda}$.

Then we have
\begin{align*}
&||{\bb\lambda}(t+1)-{\bb\lambda^*}||_2^2=\left|\left|\bb{\lambda}(t)-\alpha(t)g\transpose(t)-{\bb\lambda^*}\right|\right|_2^2\\
&=||\bb{\lambda}(t)-{\bb\lambda^*}||_2^2-2\alpha(t)g\transpose(t)\left(\bb{\lambda}(t)-{\bb\lambda^*}\right)+\alpha^2(t)\left|\left|g(t)\right|\right|_2^2\\
&\le ||\bb{\lambda}(t)-{\bb\lambda^*}||_2^2-2\alpha(t)\left(D(\bb{\lambda}(t))-D(\bb{\lambda}^*)\right)+\alpha^2(t)\left|\left|g(t)\right|\right|_2^2.\label{eq:sub_gradiant_1}
\end{align*}
Applying the inequality above recursively, it follows
\begin{align*}
& ||{\bb\lambda}(t+1)-{\bb\lambda^*}||_2^2\le||\bb{\lambda}(1)-{\bb\lambda^*}||_2^2\\
&\quad\quad-2\sum_{i=1}^t\alpha(i)\left(D(\bb{\lambda}(i))-D(\bb{\lambda}^*)\right)+\sum_{i=1}^t\alpha^2(i)\left|\left|g(i)\right|\right|_2^2.
\end{align*}
By the fact $||\bb{\lambda}(t)-{\bb\lambda^*}||_2^2\ge 0 $ and $||\bb{\lambda}(1)-{\bb\lambda^*}||_2^2\le R$, where $R$ is a predetermined bound of $\lambda$, we have
\begin{equation*}
2\sum_{i=1}^t\alpha(i)\left(D(\bb{\lambda}(i))-D(\bb{\lambda}^*)\right)\le R^2+\sum_{i=1}^t\alpha^2(i)\left|\left|g(i)\right|\right|_2^2
\end{equation*}
Combining this with
\begin{align*}
&\sum_{i=1}^t\alpha(i)\left(D(\bb{\lambda}(i))-D(\bb{\lambda}^*)\right)\\
&\quad\ge \sum_{i=1}^t\alpha(i)\min_{i=1,\dots,t}D(\bb{\lambda}(i))-D(\bb{\lambda}^*)\\
&\quad= \sum_{i=1}^t\alpha(i) D(\bb{\lambda}_{best})-D(\bb{\lambda}^*)
\end{align*}
and by the assumption $\left|\left|g(i)\right|\right|_2^2\le G^2$, we have
\begin{align}
\label{eq:sub_gradient_ineq}
0\le D(\bb{\lambda}_{best})-D(\bb{\lambda}^*)\le \frac{R^2+G^2\sum_{i=1}^{t}\alpha^2(i)}{\sum_{i=1}^{t}\alpha(i)}
\end{align}
Therefore, when $\bb{\alpha}$ is \emph{Square summable but not summable} , \ie
$\left|\left|\bb{\alpha}\right|\right|_2^2=\sum_{i=1}^t \alpha^2(i) <\infty$, and $\sum_{i=1}^t \alpha(i)= \infty$, we have the convergence results.
\end{proof}

\subsection{More simulation results for different scenarios}
The simulation setting is the same as the one in Sec.~\ref{sec:performance}. We have four difference pathloss model\footnote{Here $L_{ow}$ is the penetration loss of an outdoor wall, 20dB.}  shown in Tab.~\ref{tab:pathloss}. The first scenario in Tab.~\ref{tab:pathloss} is the one in Sec.~\ref{sec:performance}.
\begin{table*}[htb]
\centering
\begin{tabular}{|c|c|c|}
\hline
 \multicolumn{2}{|c|}{scenario}& Pathloss\\
\hline
1) Urban & UE insider the door& $\!38.46\!+\!20\log_{10}R\!+\!0.7d_\text{2D,indoor}\!+\!18.3n^{\frac{n+2}{n+1}\!-\!0.46}+q\,L_{iw}$\\
\hline
2) Urban & UE outsider the door& $\max\left(38.46+20\log_{10}R, 15.3+37.6\log_{10}R\right)+\!0.7d_\text{2D,indoor}\!+\!18.3n^{\frac{n+2}{n+1}\!-\!0.46}+q\,L_{iw}+L_{ow}$\\
\hline
3)Suburban&  UE insider the door& $38.46\!+\!20\log_{10}R\!+\!0.7d_\text{2D,indoor}\!+\!18.3n^{\frac{n+2}{n+1}\!-\!0.46}$\\
\hline
4)Suburban&  UE outsider the door& $\max\left(38.46+20\log_{10}R, 15.3+37.6\log_{10}R\right)+\!0.7d_\text{2D,indoor}\!+\!18.3n^{\frac{n+2}{n+1}\!-\!0.46}+L_{ow}$\\
\hline
\end{tabular}
\caption{Pathloss model for different indoor small cell scenario}
\label{tab:pathloss}
\end{table*}

\begin{figure}[htb]
\centering
\includegraphics[scale=0.45]{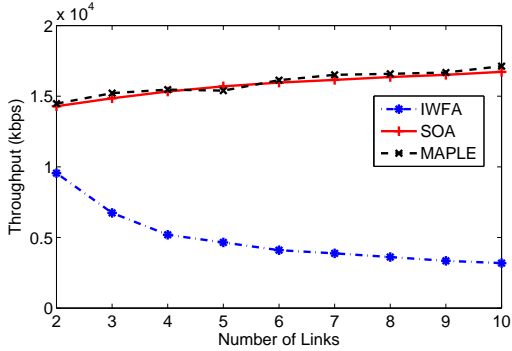}
\caption{Scenario 2) with 10m cell radius.}
\label{fig:simulation_10}
\end{figure}

\begin{figure}[htb]
\centering
\includegraphics[scale=0.45]{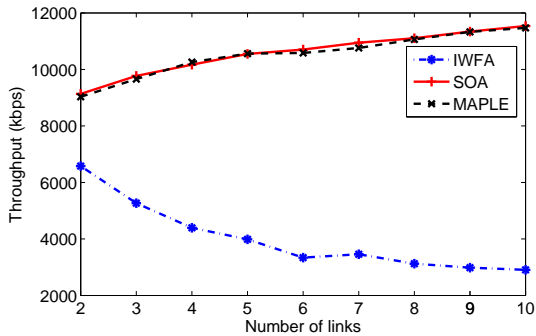}
\caption{Scenario 3) with 30m cell radius.}
\label{fig:simulation_30}
\end{figure}

\begin{figure}[htb]
\centering
\includegraphics[scale=0.48]{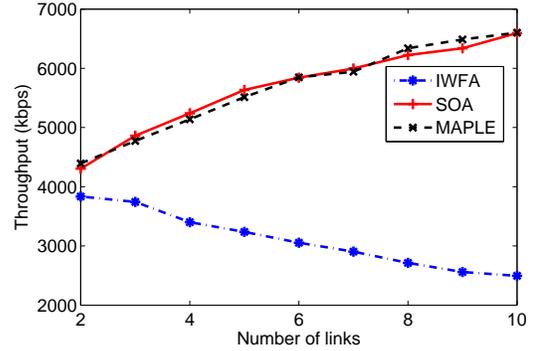}
\caption{Scenario 4) with 40m cell radius.}
\label{fig:simulation_40}
\end{figure}

For each scenario, we test the cell radius from 10m to 40m. We observe that the trends of the these curves are rather similar as Fig.~\ref{fig:simulation}, therefore three other results for different scenarios and different cell radium are picked here, as shown  in Fig.~\ref{fig:simulation_10}, Fig.~\ref{fig:simulation_30} and Fig.~\ref{fig:simulation_40}. Readers can refer to Sec.~\ref{sec:performance} for performance analysis.

\subsection{Discussions of Implementation-related Issues}
In this section, we give some discussions of implementation-related issues for the proposed framework in practical system.
\subsubsection{About Signaling Errors}
What happens if some links fail to receive the signaling from other links in the signaling slot? Does the scheduling of the resource allocation algorithms in the transmission module still work?
We discuss this problem in two cases:

\emph{Case 1:} There are  very weak interferences or almost no interference between some links, so they cannot receive signals from each other. For example, there is a long enough distance between them. Since they are just transparent to each other, there is no problem that each of them runs the proposed framework. It is actually a good case, since it can explore the spatial diversity.

\emph{Case 2:} There are substantial or strong interferences among links, but somehow some of links, say a link $A$, loses some signals from others in some channel, say a link $B$'s signal in a channel $c$. Then link $A$ would think that channel $c$ is a bad channel for link $B$ (since the channel gain is near 0). For this scenario, the scheduling and resource allocation algorithms still work, but there may be collisions in the channel allocation, and the data rate of each link decreases due to the interference. For example, both link $A$ and link $B$ choose to transmit in channel $c$. How to solve this collision issue? First, the collision is easy to detect after it happens: the data rate in the collision channel is not as high as the algorithm calculated. Once a link knows there is a collision happened in one tone, there are many ways to solve the problem. A simple one may be that the link gives up the tone with some probability and recalculates scheduling and resource allocation for the remaining channels in the next slot.

In summary, the proposed framework is robust to errors in signaling, but there will be performance loss in case 2. How much is the loss, depends on how often the error happens and how we design links to avoid  collisions when a collision happened (e.g., the probabilities to give up tones in the aforementioned mechanism). But since the proposed framework is designed for a small cell network, where links are located in a small range of (usually indoor) area, both case 1 and case 2 rarely happen.

\subsubsection{About Control Overheads}
Major control overhead comes from signaling module. How often we should do signaling depends on how fast the channels change: the slower the channel variations, the less often the signaling, and thus the less overhead. The proposed framework is designed for the small cell network scenario, which will be deployed in hot spots with low mobility where the channel quality only varies slowly with time.

Secondly, the proposed framework is able to work for general multi-channel system, since we do not have any limits on the physical layer technology. And there will also have many different implementation ways when we customize the signaling mechanism into a particular multi-channel system, since we leave many design freedoms for the signaling mechanism.
In a nutshell, the signaling scheme just requires two transmissions  with particular power levels for each link in each channel. To decrease the overhead, the length of the sub-slot can be very short, as long as that the power strength can be detected by other users in the sub-slot.

Further, we can also add channel prediction algorithms to further decrease the signaling frequency. It decreases the system overhead but may increase the estimation error. It is important to choose a proper tradeoff of frequencies of running signaling and channel perdition algorithm according to the network environment.

\end{document}